\documentclass[aps,twocolumn,groupedaddress,superscriptaddress,showpacs]{revtex4-1}

\usepackage{amssymb,amsfonts,amsmath}

\usepackage{graphicx}

\usepackage{mathptmx}

\begin{document}
\title{Overspill avalanching in a dense reservoir network}

\author{George L. Mamede}
  \affiliation{University of the International Integration of the African-Brazilian Portuguese Speaking Countries - UNILAB, Campus do Pici, 60455-760 Fortaleza, Brazil}

\author{Nuno A. M. Ara\'ujo}
   \email{nuno@ethz.ch}
   \affiliation{Computational Physics for Engineering Materials, IfB, ETH Zurich, Schafmattstrasse 6, CH-8093 Zurich, Switzerland}

\author{Christian M. Schneider}
   \affiliation{Computational Physics for Engineering Materials, IfB, ETH Zurich, Schafmattstrasse 6, CH-8093 Zurich, Switzerland}
   \affiliation{Department of Civil and Environmental Engineering, MIT, 77 Massachusetts Avenue, Cambridge, MA 02139, USA}

\author{Jos\'e C. de Ara\'ujo}
   \affiliation{Departamento de Engenharia Agr\'icola, Universidade Federal do Cear\'a, Campus do Pici, 60455-970 Fortaleza, Brazil}

\author{Hans J. Herrmann}
   \affiliation{Computational Physics for Engineering Materials, IfB, ETH Zurich, Schafmattstrasse 6, CH-8093 Zurich, Switzerland}
   \affiliation{Departamento de F\'isica, Universidade Federal do Cear\'a, Campus do Pici, 60451-970 Fortaleza, Cear\'a, Brazil}

\begin{abstract}
Sustainability of communities, agriculture, and industry is strongly
dependent on an effective storage and supply of water resources. In some
regions the economic growth has led to a level of water demand which can
only be accomplished through efficient reservoir networks. Such
infrastructures are not always planned at larger scale but rather made
by farmers according to their local needs of irrigation during droughts.
Based on extensive data from the upper Jaguaribe basin, one of the
world's largest system of reservoirs, located in the Brazilian semiarid
northeast, we reveal that surprisingly it self-organizes into a
scale-free network exhibiting also a power-law in the distribution of
the lakes and avalanches of discharges. With a new
self-organized-criticality-type model we manage to explain the novel
critical exponents. Implementing a flow model we are able to reproduce
the measured overspill evolution providing a tool for catastrophe
mitigation and future planning.
\end{abstract}

\maketitle

Water is simultaneously the most relevant commodity for life
and the primary medium through which people and the environment are
affected by natural disasters
\cite{Tilman99,Jury05,IPCC,Parry07,Ward08,Piao07}. According to the Food
and Agriculture Organization of the United Nations, the water demand in
the last century has grown more than twice the population increase and
as forecast, by $2050$, about $2\mbox{ billion}$ people will be living
in water scarcity \cite{FAO}. This is not only related to the
availability of freshwater but also with the lack of proper storage and
distribution infrastructures. It is thus of paramount interest to
understand how to guarantee an even spatial and temporal distribution of
water, avoiding the imbalance between supply and demand. The first step
in this direction, which we address here, is to analyze how the
interplay between the rainfall and evaporation affects the filling level
in the reservoirs and the network.

In semiarid environments like the Brazilian northeast it is common to
have droughts drastically affecting the crops. The weather in the region
is characterized by intermittent rains and long periods of water
scarcity. The water supply, in the dry season, is assured by a dense set
of surface reservoirs, which are essential for the sustainability of the
region, since during the rainy season water is stored to be distributed
afterwards. A proper reservoir size is crucial: while smaller reservoirs
are typically unable to supply enough water throughout the entire dry
season, due to the low storing capacity and high losses through
evaporation and infiltration, too large reservoirs can have serious
consequences for life sustainability.  For example, if the storage
capacity in a basin exceeds the volume equivalent to three times the
average annual runoff, its water will not be sufficiently renewed
\cite{Malveira12}. Not only the properties of each reservoir are
relevant but also the way they are interconnected. For example, upstream
reservoirs may retain a significant part of the collected water which
enables an efficient spatial distribution by ensuring a continuous
connection with other reservoirs even during dry seasons, when usually
no water flows through the river network
\cite{Mamede08,Trush00,LimaNeto11}. The available data from
interferometric radar and weather stations
\cite{Alsdorf00,Alsdorf07,Lincoln07,Syed10} provide the necessary
ingredients for a systematic analysis of the existing network, which is
fundamental to further planning for improvement. We introduce here a
hydrological model for the evolution of water resources in the network
and apply it to the upper Jaguaribe basin, a semiarid region in the
Brazilian northeast, for which we have collected information about the
rainfall and the distribution of reservoirs over two decades. 

\section{The Reservoir Network}

\begin{figure*}
\begin{center}
\includegraphics[width=0.8\textwidth]{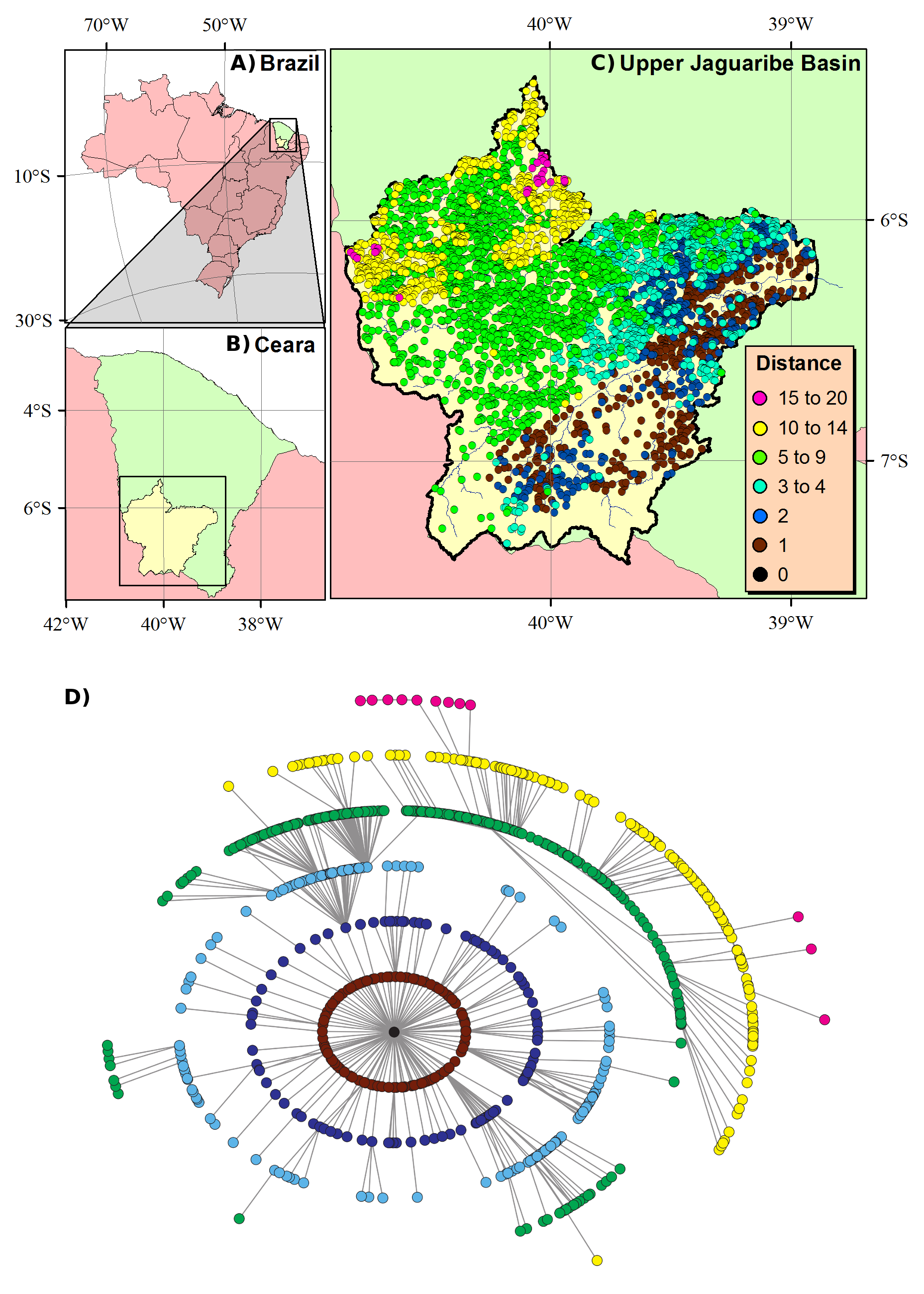}\\
\end{center}
\caption{
 Geographical location of the Or\'os Basin, in the Brazilian northeast
(A). Located in the state of Cear\'a (B), the Upper Jaguaribe Basin
consists of $3\mbox{ }978$ water reservoirs with a total catchment area of about
$25\mbox{ }000 \mbox{ km}^2$. The reservoirs have been identified through
satellite imagery from three wet years, namely, 2004, 2008, and 2009.
In (C) we mark the reservoirs where the color scheme represents the
distance to the outlet in units of intermediate reservoirs. The network
can be schematized by a graph (D) where nodes represent the reservoirs
and links stand for the connections between them. For the sake of
graphical clearness, reservoirs without incoming links have not been
included. \label{fig::fig1}}
\end{figure*}

Located in the state of Cear\'a, northeast of Brazil, the upper
Jaguaribe basin has a catchment area of about $25\ 000~\mbox{ km}^2$ and
almost $4\ 000$ reservoirs (see Fig.~\ref{fig::fig1}). Characterized by
a broad reservoir-size distribution, ranging from $2\ 500~\mbox{ m}^3$
to $2~\mbox{ billion m}^3$, this basin stores water for more than
half-million people. It has an average annual rainfall of $860~\mbox{
mm}$ and a potential evaporation as high as $2\ 000~\mbox{ mm}$. The
weather conditions together with shallow soils above crystalline
bedrocks make the rivers intermittent and leads to runoff rates below
$7\%$ of the precipitation. The main river is the Jaguaribe, which
contributes to the Or\'os reservoir located at the basin outlet, the
black reservoir in Fig.~\ref{fig::fig1} (northeast). The latter, with a
storage capacity of $1\ 940\mbox{ million m}^3$, is the second largest
reservoir of the state and an important water reserve, which yields
about $630~\mbox{ million m}^3/\mbox{year}$ with an annual reliability
level of $90\%$, i.e., $90\%$ probability of being able to supply this
amount every year.

\begin{figure}
\includegraphics[width=\columnwidth]{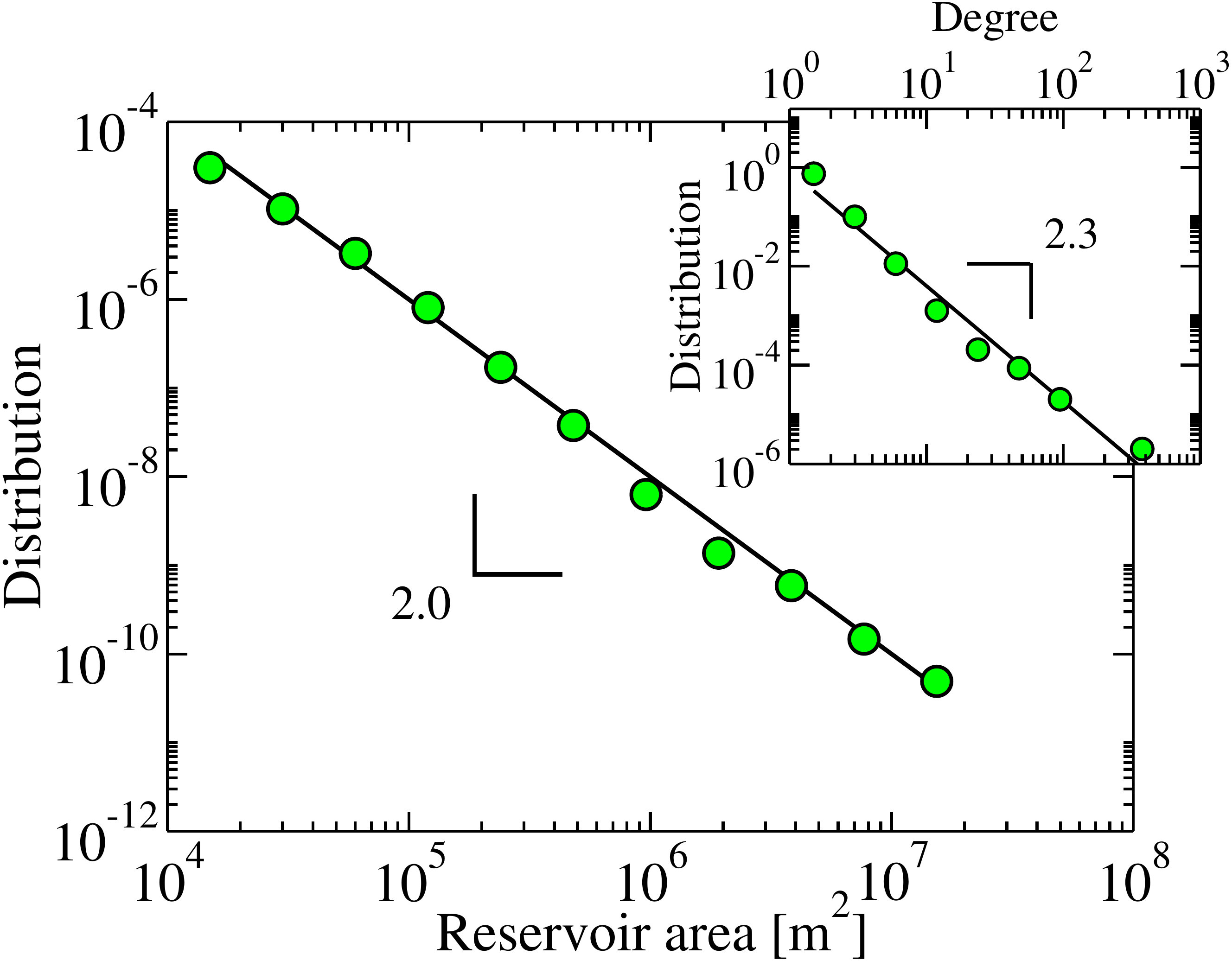}
\caption{
 The reservoirs in the upper Jaguaribe basin are characterized by a
broad distribution of areas and connections. \textbf{Main plot:} The
distribution of areas $P_A(A)$ scales according to a power law,
$P_A(A)\sim A^{-\lambda}$, with $\lambda=2.00\pm0.05$. \textbf{Inset:}
The degree distribution $P_k(k)$ of the current reservoir network
follows a power law, $P_k(k)\sim k^{-\gamma}$, with $\gamma=2.3\pm0.2$.
The size and location of the reservoirs were measured from satellite
images (see section \textit{Materials and Methods}). \label{fig::fig2}}
\end{figure}

Analyzing satellite images from the wet years (see section
\textit{Materials and Methods}), we have identified the geographical
location and size of each reservoir, as well as the connection between
them. The location of the full set of reservoirs in the Jaguaribe basin
is represented in Fig.~\ref{fig::fig1}~C. The color scheme stands for
the distance from the outlet (Or\'os), in units of reservoirs. As shown
in Fig.~\ref{fig::fig1}~D, to characterize the network we schematize it
as a graph where reservoirs are nodes and the connections between them
are links. The links are directional since flow goes from upstream to
downstream, based on the height difference between reservoirs. The
collected data disclose a network of reservoirs organized in a tree with
low branching rate. All reservoirs have a single outlet connection but
with a broad distribution of inlets. While most upstream reservoirs have
no inlet connection, for the ones downstream typically multiple
connections exist (see Fig.~\ref{fig::fig1}~D). The degree $k$ of the
reservoir (i.e., number of connections) varies from one to around $400$
and the degree distribution $P_k(k)$, is a power law, $P_k(k)\sim
k^{-\gamma}$, with a degree exponent $\gamma=2.3\pm0.2$, as shown in the
inset of Fig.~\ref{fig::fig2}.

The satellite images from the wet years allow to measure the area ($A$)
of each reservoir disclosing a power-law distribution of areas,
$P_A(A)\sim A^{-\lambda}$, with $\lambda=2.00\pm0.05$ (see
Fig.~\ref{fig::fig2}). From the area, the storage capacity ($V$) can
also be estimated. Molle \cite{Molle89} has studied $417$ reservoirs of
the Brazilian semiarid region, a $970\ 000~\mbox{ km}^2$ domain in the
northeast and southeast of Brazil, and proposed an area/volume relation
given by,
\begin{equation}\label{eq::areavolume}
A=cd\left(\frac{V}{d}\right)^{(c-1)/c}, \end{equation}
where $c$ and $d$ have been established empirically to be $c=2.7$ and
$d=1500$. For the upper Jaguaribe basin, Mamede \cite{Mamede08} tested
this equation for $21$ reservoirs showing satisfactory agreement. Since
the capacity of a reservoir cannot be directly measured with the
satellite imagery, we apply the same relation (Eq.~\ref{eq::areavolume})
to the remaining reservoirs. 

\section{Hydrological Model}

The planning of an efficient reservoir network requires the analysis of
its response to typical weather conditions. The amount of water in a
reservoir is a dynamical quantity, with a positive contribution from
precipitation and upstream inflow, while mechanisms such as groundwater
infiltration, evaporation, and overflow tend to reduce it. To quantify
the precipitation and evaporation, meteorological time series for
monthly evaporation level and rainfall (volume per area) have been
obtained for the $131$ weather stations of the region \cite{Funceme}.
From the topology of the basin, the watershed lines have been traced and
the subbasin of each reservoir determined. The daily average of rainfall
and evaporation in the subbasin is determined from the closest station.
The net evaporation is estimated based on the mean reservoir area, while
the contribution of the precipitation to the reservoir level accounts
for both direct rainfall in its area and a fraction of the one falling
in the area of the subbasin drained into it.  The latter fraction is
denoted as runoff and depends on factors such as soil constituents,
vegetation coverage, rain intensity (both actual and preceding), and
topography \cite{Easterling07,Barnett09,Koster10,Pagano10}. For
simplicity, we take the same value of the runoff ($0.04$) for the entire
region, i.e., $4\%$ of the rainfall in the subbasin, outside the
reservoir, is considered to drain into the reservoir. As discussed
below, a good agreement is found for the time evolution of the water
volume in different reservoirs with this uniform runoff coefficient.

To model the time evolution of the water resources in the reservoir
network we consider that initially each reservoir is at $20\%$ of its
storage capacity (presented results are not affected by the initial
conditions). Following the time series for the rainfall and evaporation,
the inflow of water to each reservoir is computed on a daily basis
accounting for three contributions: direct rainfall, water drained from
the subbasin, and inflow from reservoirs upstream.  Reservoirs are
updated from upstream to downstream and the overflow is obtained from
the difference between the total amount of water and the storage
capacity. The exceeding water is transferred to the reservoirs
downstream, where we neglect transmission losses through the river
transport. When an overflow occurs, water is transferred to the next
reservoir which can trigger a cascade of overflows of other reservoirs
downstream. We denote as avalanche this type of events and we measure
their size as the number of connected reservoirs which overflow within
the same day. 

\section{Water Transport}

For simplicity, let us start by considering a uniform distribution of
rainfall and evaporation over the entire region during two decades. At
the subbasin level, a competition takes place between the increase of
water due to the rain and upstream inflow and the evaporation.
Reservoirs where the influx is larger than the outflux, after some
initial filling time, are full and constantly overflowing, i.e.,
transferring water to the one downstream. On the other hand, reservoirs
where losses are not compensated by incoming water, essentially dry out
and overflow never occurs. As an example, we take a level of
precipitation of $1.825~\mbox{ mm}$ and an evaporation of $6.832~\mbox{
mm}$ (per unit area and day), corresponding to the daily averages for
the period between $1991$ and $2010$. When we initialize the system with
all reservoirs at $20\%$ of their capacity, after $15$ years, every day,
one avalanche affecting $1258$ reservoirs occurs. This scenario is not
observed in reality since both the precipitation and evaporation are
asymmetrically distributed in space and time. Therefore, a meaningful
modeling of the water transport requires a proper quantitative
description of these two key processes.

\begin{figure}
\includegraphics[width=\columnwidth]{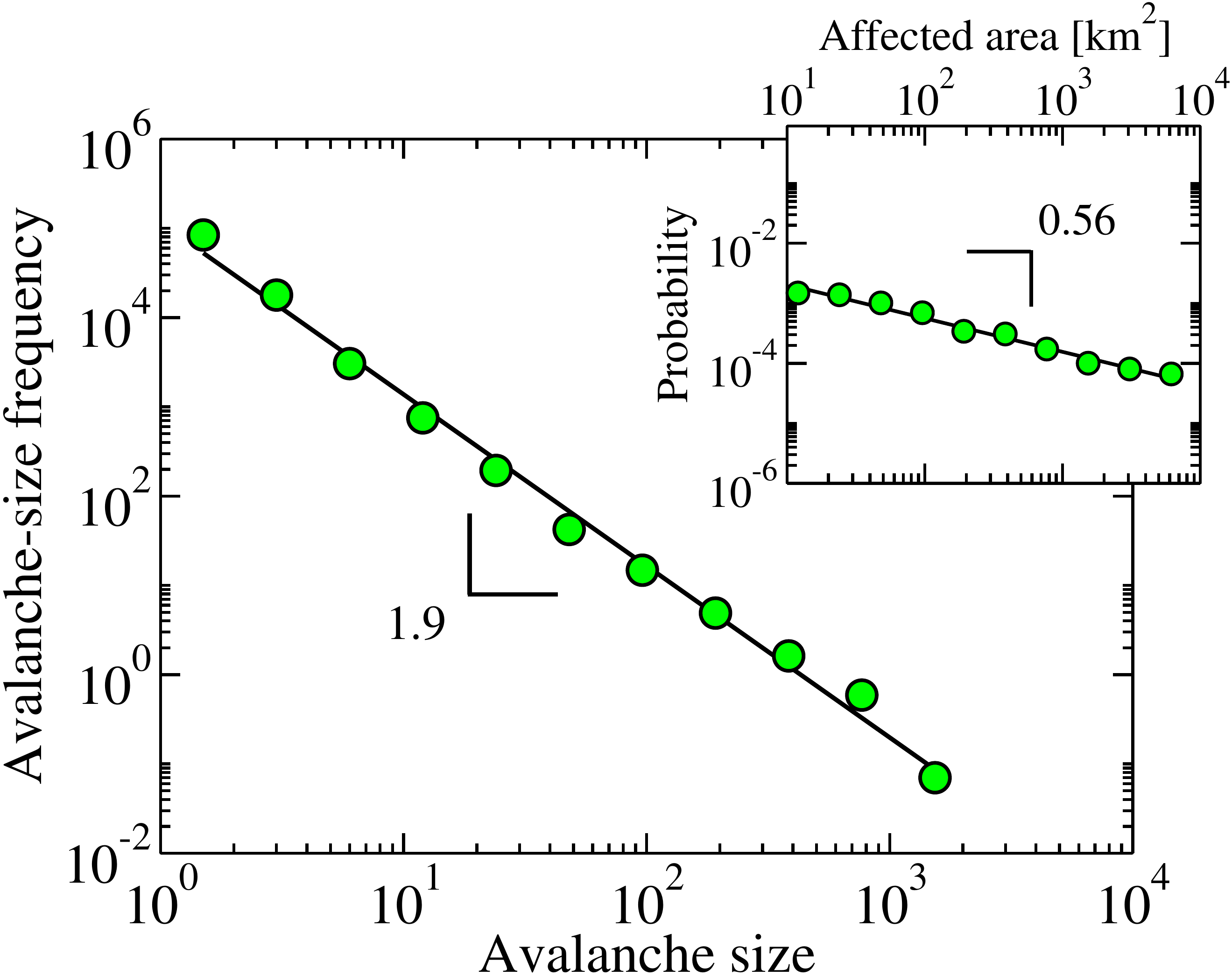}
\caption{
 Power laws over more than three orders of magnitude, revealing a lack
of a characteristic size on the avalanches and affected areas.
\textbf{Main plot:} The avalanche-size histogram scales as $H(s)\sim
s^{-\alpha}$, with $\alpha=1.9\pm0.2$, where the avalanche size $s$ is
defined as the number of connected reservoirs with overflow.
\textbf{Inset:} The distribution of affected area $P_{A_f}(A_f)\sim
A_f^{-\beta}$, with $\beta=0.56\pm0.08$. The area $A_f$ corresponds to
the total area of catchments related to the avalanche. Both results have
been numerically obtained for a runoff coefficient of $0.04$ and with
the real time series for rainfall and evaporation, between $1991$ and
the end of $2010$.  \label{fig::fig3}}
\end{figure}

Inserting the time series for rainfall and evaporation from $1991$ till
the end of $2010$, we have followed the dynamics of the reservoir
network and measured the size of the avalanches as well as the volume of
water storage at each reservoir. We define the avalanche size $s$ as the
number of connected reservoirs from which water overflows during the same day.
Over these two decades, in the rainy seasons, many avalanches are
observed with sizes ranging over three orders of magnitude. Most of the
avalanches solely affect two reservoirs, but large ones also occur
(see \textit{Supporting Video}). The avalanche-size histogram $H(s)$,
shown in Fig.~\ref{fig::fig3}, scales with the avalanche size as
$H(s)\sim s^{-\alpha}$, with $\alpha=1.9\pm0.2$. The power law behavior
is related to the absence of a characteristic scale in the
avalanches (see section \textit{Stochastic Model}). In the considered
time interval, the largest observed avalanche occurred during the day
$47$th of $2004$ and involved about $2000$ reservoirs, i.e., more than
$50\%$ of the entire set. In the inset of Fig.~\ref{fig::fig3} we also
show the distribution of the area $P_{A_f}$ affected during an avalanche.
The affected area $A_f$ is defined as the total area of catchments
related to the overflowing reservoirs. This distribution scales as
$P_{A_f}\sim A_f^{-\beta}$, with $\beta=0.56\pm0.08$. This low value of
$\beta$ reveals that large scale events evolving extended catchment
areas also occur.

\begin{figure}
\includegraphics[width=\columnwidth]{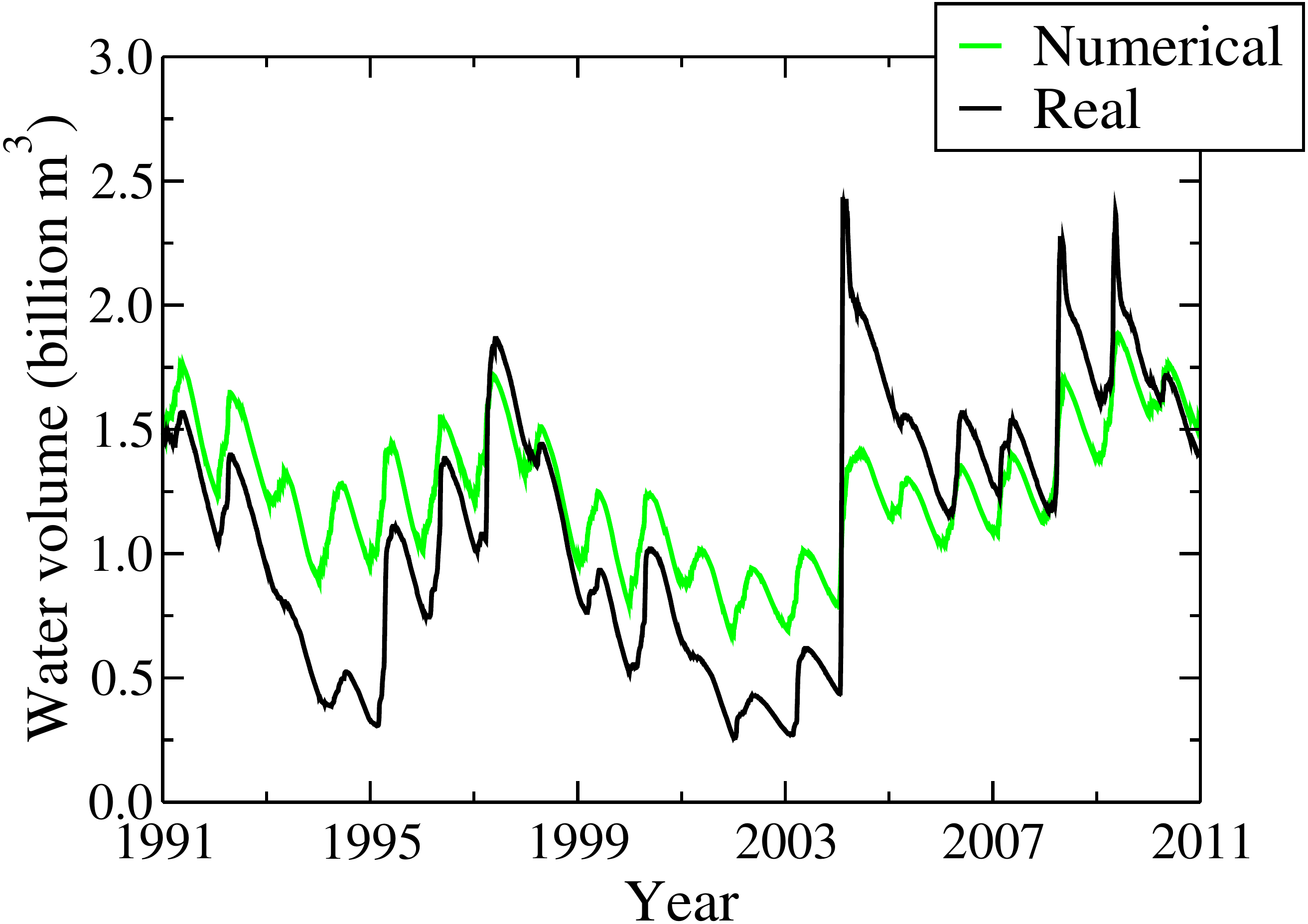}
\caption{ 
 Model comparison with empirical results for the temporal evolution of the
water volume at the Or\'{o}s reservoir, revealing a good quantitative
agreement between both. Numerical results have been obtained with
simulations for a runoff coefficient of $0.04$ and the real data
corresponds to the time series of the water volume in the interval
between $1991$ and $2011$ \cite{Cogerh}. The larger differences are
observed after long periods of intensive rain (like the beginning of
$2004$) since the wet soils improve the amount of water drained from the
subbasin to the reservoir. The differences can also be due to the
construction of new reservoirs during this two decades. \label{fig::fig4}}
\end{figure}

A constant runoff of $0.04$ has been considered. The exponent of the
avalanche-size histogram is resilient over a wide range of runoff
coefficients (below $0.05$). For large values of runoff some reservoirs
have a large inflow of water when compared to the outflow and, once
full, they overflow every day. While the exponent of the avalanche-size
histogram does not depend on the runoff coefficient, to reproduce the
measured water volume of each reservoir, a proper value is required. As
pictured in Fig.~\ref{fig::fig4}, with the considered runoff, the water
volume is satisfactorily reproduced over most of the considered time
range. Slight differences between the simulation and real values occur
after long periods of intensive rain. As discussed by Koster {\it et
al.} \cite{Koster10}, in this case the runoff coefficient could have
been affected by the antecedent rainfall. As observed for $2004$, after
long periods of intensive rain, the wet soils enhance the runoff,
promoting the drainage of a larger volume of water from the subbasin to
the reservoir and thus larger avalanches.

\section{Stochastic Model}

The most striking result unveiled by our hydrological model is the
power-law avalanche-size distribution. This result resembles the
fingerprint of self-organized critical (SOC) systems disclosed by P. Bak
\cite{Bak96} though with a different exponent than SOC on scale-free trees
\cite{Goh03,Luque08}. SOC is characterized by branching, where the
discharge into neighbors might trigger also their discharge under slow
driving. However, none of these features appears in the reservoir
networks. Rain falls synchronously on several
reservoirs and, while the inflow connections are scale free, the outflow
occurs through one single connection (no branching). Furthermore, water
coming from the synchronous overspill at two (or more) upstream
reservoirs joins on downstream ones and increases the amount of water
transported through a single branch.

\begin{figure}
\includegraphics[width=\columnwidth]{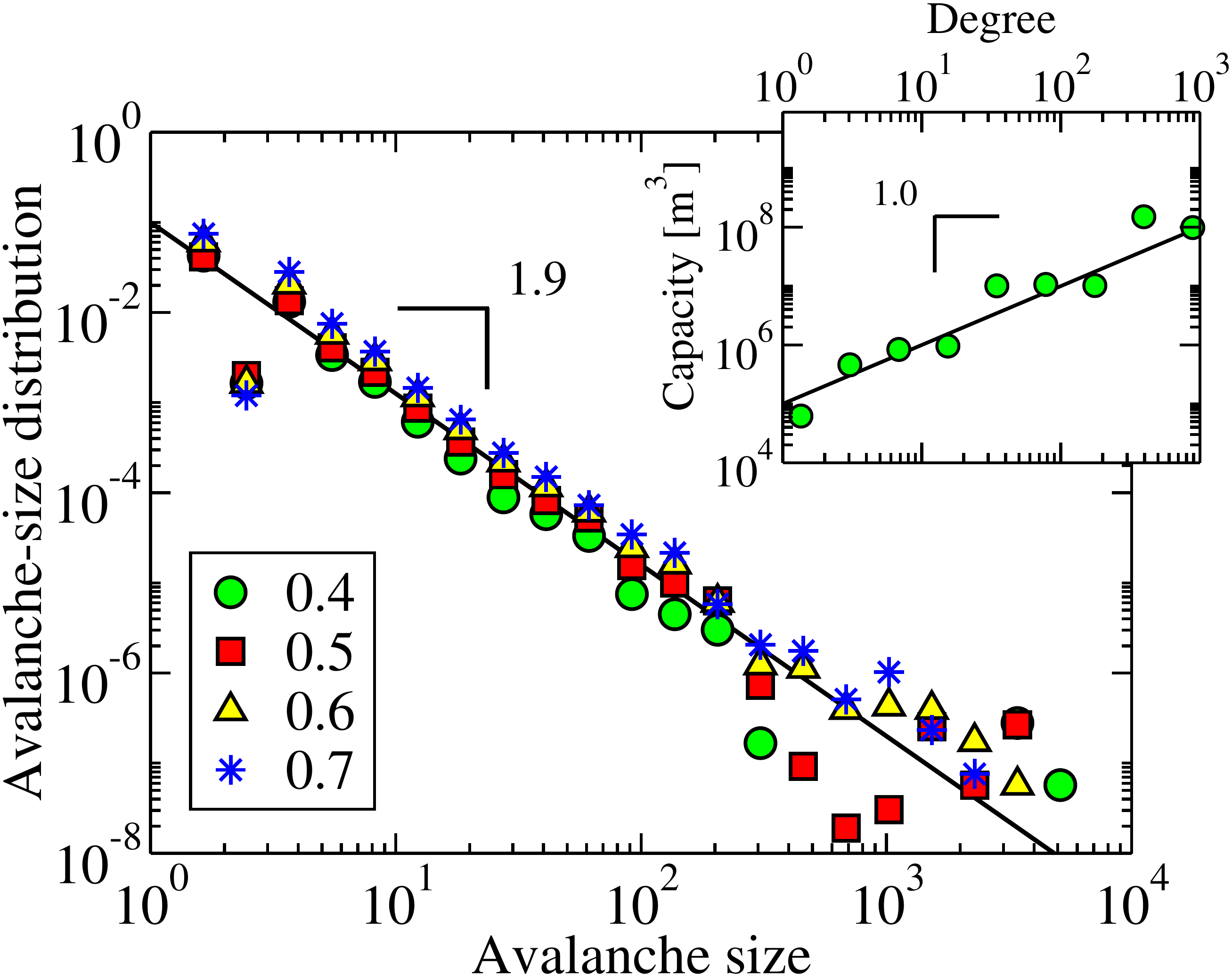}
\caption{
 Power laws with the same exponent than the ones for the hydrological
model are observed with a simplified model on the reservoir network from
the upper Jaguaribe basin. \textbf{Main plot:} Avalanche-size
distribution obtained for different values of dissipation, namely,
$0.4$, $0.5$, $0.6$, and $0.7$. Results have been averaged over $10^2$
samples of $10^5$ iterations each. The solid line stands for a power law with 
exponent $1.9$ as in the main plot of Fig.~\ref{fig::fig2}. \textbf{Inset:}
Dependence of the average reservoir capacity on the degree of the
reservoir, where the solid line corresponds to a linear relation. The
capacity of each reservoir have been estimated from satellite imagery
(see section \textit{The Reservoir Network}).
 \label{fig::fig5}
}
\end{figure}

Let us consider the following model for the same set of $N$ reservoirs from
the upper Jaguaribe basin. Each reservoir $i$ is characterized by its
amount of water $m_i$. Iteratively, a unit volume of water is added to
$N$ randomly selected reservoirs. Though on average each reservoir
receives the same amount of water, at iteration $i$ a reservoir can be
selected more than once. To each reservoir a threshold $t_i$ is assigned
such that if the amount of water in the reservoir exceeds this
threshold, $m_i>t_i$, overspill occurs and a fraction $1-f$ of water is
transferred to the downstream reservoir while the remaining water $f$ is
dissipated (evaporation and transmission losses). Since the inset of
Fig.~\ref{fig::fig5} shows a linear relation between the capacity of a
reservoir and its number of connections (degree), we take $t_i$ as equal
to the degree. In the main plot of Fig.~\ref{fig::fig5} we show the
avalanche-size distributions for different values of dissipation $f$
obtained simulating the above model on the $3\mbox{ }978$ reservoirs
identified in the upper Jaguaribe basin.  For a wide range of $f$,
results are consistent with a power-law distribution extended over three
orders of magnitude with the same exponent $\alpha=1.9$ obtained with
the hydrological model in Fig.~\ref{fig::fig3} (solid line).

\section{Final Remarks}

In summary, we found some unexpected scale-free behavior in the dynamics
of water transport on a reservoir network, using available data for the
geographical distribution of reservoirs as well as the time series of
rainfall and water evaporation. The hydrological model for the in- and
outflow balance at each reservoir reveals an excellent qualitative and,
in most of the cases, quantitative agreement for the upper Jaguaribe
basin over two decades. The analysis of these reservoirs, with satellite
imagery, discloses a tree-like scale-free network with a power-law
distribution of capacities. Our model for the water transport shows a
cascade dynamics of connected spilling over reservoirs characterized by
overflows extended over three orders of magnitude in length scales. The
introduced stochastic model grasps the main features observed with the
hydrological model. With this scheme it is now possible to characterize
the role of the network topology, spatial and time correlation of the
rainfall, and evaporation, on water flow. The proposed tools aim also to
help the planning of new reservoirs by numerically testing their impact
under typical weather conditions, avoiding either scarcity over dry
seasons or floods in rainy ones.  Despite the degree of agreement
between numerical and real data, further improvement of the model may
reproduce even more details.  Future work can account for the temporal
evolution of the network itself and its impact in the region
\cite{Poff07}. In this study the water transport between reservoirs has
been considered to occur within one day and without transmission losses,
a more detailed description of the mass transport could also be included
\cite{Keer07,Moyle07}. Upstream reservoirs also serve as sediment
detention basins, retaining a considerable amount of sediment generated
within the catchment and extending the lifetime of larger reservoirs
located downstream \cite{Mamede08,LimaNeto11}, affecting the storage
capacity and, consequently, the entire dynamics
\cite{Araujo06,Malmon05}.  Understanding the impact of such mechanism
may also be an interesting focus of research.

\section{Materials and methods}

To identify the reservoirs in the network and quantify their maximum
capacity we analyzed images from three wet years, namely, 2004, 2008,
and 2009. These images, collected by the satellites LANDSAT (1, 2, 3, 5,
and 7) and CBERS (2 and 2B), were obtained from the website of the
``Instituto Nacional de Pesquisas Espaciais'' \cite{INPE}. The LANDSAT
generates images in seven bandwidths: three in the visible region, one
in the near infrared, two in the mid-infrared, one in the thermal
infrared, and a panchromatic one. By properly selecting reference
points, the obtained images have been georeferenced and prepared to be
used in the Geographical Information System (GIS). Through unsupervised
classification techniques, which do not require any further information
besides the images, the data have been processed and pixels grouped by
spectral features to identify the ones corresponding to water
reservoirs. An additional handmade filtering was necessary to avoid
misleading information due to the presence of clouds. By superposing the
images for different years, it was possible to locate each reservoir and
trace its perimeter, obtaining the maximum area. With the relation
proposed by Molle \cite{Molle89}, Eq.~\ref{eq::areavolume}, the maximum
capacity of each reservoir has been estimated. The drainage network was
obtained from the analysis of high-resolution elevation data, collected
by the Shuttle Radar Topography Mission (SRTM) from NASA
\cite{Farr07,SRTM}, by tracing the watershed lines, and afterward each
reservoir was located in this network. The dataset with the network and
geographical localization of each reservoir is available as
\textit{Supporting Information}. The daily average of rainfall for each
reservoir is obtained from the closest weather stations and the
evaporation is estimated from the monthly average over several decades
\cite{Funceme}. The pan evaporation data accounts for the evaporation on
the surface of the reservoir (about $70\%$ of class A pan evaporation)
\cite{Linacre94}, for the infiltration, as well as for the water
extracted for several uses (about $30\%$) \cite{Molle89,Malveira12}. 

\begin{acknowledgments}
We acknowledge financial support from the ETH Risk Center. We also
acknowledge the Brazilian agencies CNPq, CAPES and FUNCAP, and the
Pronex grant CNPq/FUNCAP, for financial support. This study has mainly
been supported by a grant of the ETH North-South Centre. We thank COGERH
-- Water Management Agency for providing data about several reservoirs.
\end{acknowledgments}

\end{document}